\documentclass[twocolumn]{revtex4-1}

\def\gs{\Upsilon_\star}
\def\sun{\odot}

\def\be{\begin{equation}}
\def\ee{\end{equation}}

\def\la{\label}
\def\bea{\begin{eqnarray}}
\def\eea{\end{eqnarray}}

\def\ci{\cite}
\def\la{\label}

\def\rrc{r_{c}}

\usepackage{ctable}
\usepackage{multirow}
\usepackage{graphicx}
\usepackage{grffile}
\usepackage{epstopdf}
\usepackage{subfig}
\usepackage{subfloat}
\usepackage{rotating}

\begin{document}
\title {Galactic Phase Transition at  $E_c= 0.11\,eV$   from  Rotation Curves
of Cored LSB\\
and  nonperturbative Dark Matter Mass}
\author {Axel de la Macorra}
 \author {Jorge Mastache}
\affiliation{Instituto de Fisica, Universidad Nacional Autonoma de Mexico, Ciudad Universitaria, 01000,  D.F., Mexico}
\author{Jorge L. Cervantes-Cota}
\affiliation{Depto. de Fisica, Instituto Nacional de Investigaciones Nucleares,  Apdo. Postal 18-1027, 11801, D.F. Mexico}

\begin{abstract}
We analyze the a set of seventeen rotation curves of Low Surface Brightness (LSB) galaxies from the The HI Nearby Galaxy
Survey (THINGS) with different mass models to study the core structure and to determine a phase transition energy
scale ($E_c$) between hot and cold dark matter,  due to nonperturbative  effects in the Bound Dark Matter (BDM) model.   Our results agree with previous ones implying the cored profiles are preferred over the N-body motivated cuspy
Navarro-Frenk-White (NFW) profile.  We find an average galactic core radius of
$r_c = 260\times 10^{\pm 1.3}$ pc and a phase transition energy $E_c = 0.11\times 10^{\pm  0.46}{\rm \ eV}$, that is of the same order of magnitude as the sum of the neutrino masses.

\end{abstract}


\maketitle

\textbf{Introduction.}
Understanding the distribution of dark matter (DM) in galaxies has been a major work in recent
times \cite{KrAl78, deBlok10}. Late-type Low Surface Galaxies (LSB) galaxies are of special interest since it is believed that they are dominated by DM, and high resolution $HI$, $H_{\alpha}$ and optical data can help to distinguish
among the different DM profiles proposed in the literature.  There are essentially two types of halo profiles, the ones stemming from cosmological $N$-body simulations that have a cusp in its inner region, e.g. Navarro-Frenk-White (NFW) profile \ci{Navarro:1995iw}. On the other hand, the phenomenological motivated cored profiles,  such as  the Burkert or Pseudo-Isothermal (ISO) profiles \cite{Bu95}. Cuspy and cored profiles can both fit most LSB rotation curves,   such as \ci{deBlok:2008wp,vandenBosch:1999ka}, where LSB rotation curves has been analyzed with the most relevant profiles, showing a preference for a cored inner region with constant density and the stellar mass models is relevant in determining the size of the core radius $r_c$.
Different systematics may play an important role in the observations such as noncircular motions, resolution of data and other issues \cite{vandenBosch:1999ka,  deBlok:2008wp}, as well as the type of galaxies involved \cite{EvAnWa09}.
There are attempts to reconcile both approaches through evolution of DM halo profiles including baryonic processes  \cite{SpGiHa05} to transform cuspy to shallower profiles that follow the solid-body velocity curve ($v \sim r$) observed in late-type LSB. However, this issue is a matter of recent debate \cite{Pontzen:2011}.
Recently one of us proposed a new type of Dark Matter,  called bound dark matter (BDM)  \cite{delaMacorra:2009yb}, motivated by particle physics in which DM particles  are relativistic at high energy densities, $\rho_{BDM} > \rho_c$ above the phase transition scale energy scale $\rho_c=E_c^4$, i.e. we have Hot DM (HDM) with peculiar  velocity $v \simeq c$. But, for lower energy densities than $\rho_c$ the BDM particles acquire a large mass due to nonperturbative physics and they behave as standard Cold DM (CDM) particles with $v \ll c$. The phase transition between CDM and HDM  given by the scale $\rho_c \equiv E_c^4$ can be determined theoretically or phenomenologically  by consistency with cosmological or galactic data. In the present work we will estimate its value through the study of the rotation curves of LSB galaxies.\\
The work here is twofold. On the one hand, we  use galactic rotation curves to extract information on the possible core nature of DM dominated galaxies, and on the other, we determine the fundamental parameters behind our  DM proposal.  To perform these tasks we use four different halo mass profiles (NFW, BDM, Burkert, ISO) and five stellar mass models (Min.Disk, Min.Disk+gas, Kroupa, diet-Salpeter, Free $ \Upsilon_{\star}$).  We present here the  main results for the BDM and NFW profiles, and we refer  to the most extensive work in \cite{BDM2011}. The standard deviation of the parameters, e.g. the core radius $r_c$,
depends on the stellar mass model used but the results are  consistent with each other. In some galaxies the core radius
may be different from zero at $\% 95$ confidence level and  consistent with $r_c=0$ for different stellar mass model. This
clearly shows the importance of having better inner data and a better understranding on the mass models.\\
For the analysis we use The HI Nearby Galaxy Survey (THINGS) \cite{Walter:2008wy}, which collects high resolution and excellent sensitivity of the velocity fields revealing extended measurements of gas rotation velocities and circular baryonic matter trajectories \cite{Walter:2008wy}.  We will show  that the LSB rotation curves yield a phase transition
energy scale $E_c$, between HDM and CDM for our BDM profile, at
$E_c = 0.11^{+0.21}_{-0.07} {\rm \ eV}$. This $E_c$ is a new fundamental scale for  DM and can also be theoretically determined using  gauge group dynamics. However, even though we propose $E_c$  as a new fundamental constant for DM it is important to notice that its value does depend on the choice of BDM profile used and on the quality of the observational data. The coincidence in the size of the
sum of neutrino masses with the magnitude as $E_c$  could  open an interesting  connection between the generation of DM and neutrinos masses.\\
\textbf{BDM Model.} Cosmological evolution of gauge groups, similar to QCD,  have been studied to understand the nature of dark energy \ci{Macorra.DE} and also DM \ci{Macorra.DEDM}. For an asymptotically free gauge group the strength of the fundamental interaction increases with decreasing energy and the non-perturbative  mechanism  generates the mass of bound states, as baryons in QCD.
In this case, the mass of the bound states particles is not the sum of its component particles but it is due to the binding energy  and is parameterized by $\Lambda_c$.  The condensation or phase transition scale is defined as the energy where the gauge coupling constant $g$ becomes strong, i.e. $g(\Lambda)\gg 1$, giving a condensation scale $\Lambda_c = \Lambda_i\,e^{-8\pi^2/bg^2_i}$, where $b$ is the one-loop beta function which depends only on the number of fields in the gauge group (for example for a SUSY gauge group $SU(N_c),N_f$, where $N_c(N_f)$ is the number of colors (flavors), we have  $b=3N_c-N_f$)  and $g_i$ is the value of the coupling constant at an initial scale $\Lambda_i$. Clearly,  $\Lambda_c$ is exponentially suppressed compared to $\Lambda_i$ and we can understand why $\Lambda_c$ is  much smaller then the initial $\Lambda_i$, which may be identified with the Planck, Inflation, or Unification scale. The order of magnitude of the mass of these particles is $ m_{BDM}=d\,E_c $ with $d=\mathcal{O}(1)$ a proportionality constant. In QCD one has $\Lambda_{QCD}\simeq 200\,{\rm MeV}$ with the pion (proton) mass $m_\pi\simeq140 {\rm \ MeV}$ ($m_b\simeq 939 {\rm \ MeV}$), giving  a proportionality constant is in the range $0.7<d<5$, with bound mass much larger than the mass of the quarks ($m_u\simeq (1-3) {\rm \ MeV}, m_d\simeq (3.5-6) {\rm \ MeV}$).
In our case the gauge group and elementary fields {\it are not} part of the standard model (SM). Our dark gauge group is assumed to interact with the SM only through gravity and is widely predicted by extensions of the SM, such as brane or string theories. We can relate $E_c$ to $\Lambda_c$ since the energy density depends on the average energy per particle and the particle number density $n$, i.e.   $\rho_c\equiv E_c^4 = \Lambda_c\,n$.\\
There are two natural places where one may encounter  high energy densities for dark matter. One is at early cosmological times and the second place is in galactic inner regions. Here we are interested in the latter case. Away from the center of the galaxies the energy density decreases and one has $\rho_{BDM} < \rho_c$. In this region  the BDM particles are CDM. Since CDM is well parameterized  by a cuspy NFW profile
$\rho_{NFW}=\rho_0/[r/r_s( 1+r/r_s)^2)]$ we expect BDM to have this limit away from the galactic center and as long as  $\rho_{BDM} \ll  \rho_c$. However,  in the inner region the DM energy density increases and once $\rho_{BDM}$ reaches  $\rho_c$  we have a phase transition and the BDM particles become relativistic with a dispersion velocity $v \simeq  c$ and  therefore forming a core inner region. At this stage
the NFW profile would no longer describe the behavior of our BDM. Therefore, since our BDM behaves as CDM away from the galactic center but the density has a core inner region, the proposed BDM profile is  \cite{delaMacorra:2009yb}
\captionsetup{margin=10pt,font=small,labelfont=bf,justification=centerlast,indention=-.3cm}
\begin{table}[h]
  \centering
  \scriptsize{
    \begin{tabular}{l|p{4mm}p{4mm}p{4mm}p{4mm}p{4mm}|p{4mm}p{4mm}p{4mm}p{4mm}p{4mm}|p{4mm}p{4mm}p{4mm}p{4mm}p{4mm}}
    \multicolumn{ 1}{l|}{BDM} &                                  \multicolumn{ 5}{c|}{Min.Disk } &                            \multicolumn{ 5}{c|}{Min.Disk+Gas } &                                    \multicolumn{ 5}{c}{Kroupa } \\
    \multicolumn{ 1}{l|}{Gal} &     $r_s$ & $\log \rho_o$  &     $\;\;r_c$ &     $E_c$ & $\chi^2_{r}$ &     $r_s$ &  $\log \rho_o$  &     $\;\;r_c$ &       $E_c$ & $\chi^2_{r}$ &     $r_s$ & $\log \rho_o$ &   $\;\;r_c$ &       $E_c$ & $\chi^2_{r}$  \\ \hline
            A1 &       3.6 &       7.3 &       1.3 &       .05 &       0.4 &       4.8 &       7.0 &       0.7 &       .05 &       0.3 &       4.4 &       7.1 &       1.0&       .05 &       0.2 \\
            A2 &       4.6 &       8.6 &      .001 &       1.3 &       0.6 &       4.5 &       8.6 &       0.0 &         -- &       0.5 &       6.4 &       8.2 &       0 &         -- &       1.1 \\
            A3 &       1.5 &       9.4 &       .19 &       0.2 &       4.2 &       0.9 &      10 &       0.9 &       0.1 &       4.5 &       6.4 &       7.7 &       0 &         -- &       5.1 \\
            A4 &       7.7 &        7.5 &       .01 &       0.2 &       2.0 &       6.2 &       7.7 &       0.1 &       0.1 &       1.6 &      18 &       6.7 &       0 &         -- &       1.2 \\
            A5 &       0.2 &      10.7 &       .05 &       0.4 &       1.7 &       0.1 &      12 &       0.4 &       0.2 &       1.7 &       .03 &      13.0 &       .03 &       0.8 &       1.3 \\
            A6 &       5.9 &       8.0 &       .12 &       0.2 &       1.4 &       5.4 &       8.1 &       0.1 &       0.2 &       1.3 &      23 &       6.7 &       .0 &         -- &       1.4 \\
            A7 &       7.2 &       7.4 &       .06 &       0.2 &       3.7 &       6.2 &       7.5 &       0.1 &       .15 &       3.4 &      13 &       6.9 &      0.04 &       .14 &       3.9 \\
            B1 &      17.2 &          7.0 &      18.3 &       .04 &    0.4 &      14 &       7.0 &      13.6 &       .03 &       0.4 &      46 &       6.2 &      16 &       .03 &       0.9 \\
            B2 &       2.2 &       7.9 &       2.2 &       .03 &       2.1 &       2.0 &       7.9 &       2.0 &       .02 &       1.8 &       2.0 &       7.9 &       2.04 &       .04 &       1.7 \\
            B3 &        1.9 &       9.2 &        1.9 &       .08 &       1.7 &       1.8 &       9.3 &       1.8 &       .08 &       1.6 &       2.4 &       8.9 &       1.7 &       .08 &       1.4 \\
            B4 &       2.5 &        8.5 &       2.5 &       .09 &       0.7 &       3.0 &       8.2 &       1.6 &       .09 &       0.9 &      50 &       7.7 &      47 &       .08 &       1.2 \\
            B5 &       3.7 &       8.4 &       3.7 &        0.1 &       0.6 &       3.3 &       8.5 &       3.3 &       0.1 &       0.7 &       5.8 &       7.9 &       5.8 &       0.1 &       0.4 \\
            B6 &          2.0 &       9.3 &    2.0 &       .08 &       1.4 &       1.6 &       9.5 &       2.4 &       0.1 &       3.0 &      20 &       6.7 &       0.8 &       0.1 &       1.9 \\
            B7 &      10.3 &       7.5 &      12 &       .03 &       0.3 &       9.6 &       7.5 &       9.6 &       .03 &       0.5 &      50 &       6.8 &      51 &       0.1 &       1.3 \\
    \end{tabular}
    }
  \caption{\footnotesize{BDM parameters for different galaxies and mass model with $r_s, r_c$ in kpc, $\rho_0$  in $M_\sun/kpc^3$ and  $E_c$  in eV.}}
  \label{tab:onlydm_bdm}
\end{table}
\begin{table}[h]
  \centering
  \scriptsize{
      \begin{tabular}{l|rrr|rrr|rrr}
    \multicolumn{ 1}{c|}{NFW} &       \multicolumn{ 3}{c|}{Min. Disk } & \multicolumn{ 3}{c|}{Min. Disk + Gas } &          \multicolumn{ 3}{c}{Kroupa } \\
    \multicolumn{ 1}{c|}{Gal} &     $r_s$ & $\log\rho_0$  & $\chi^2_{r}$ &     $r_s$ & $\log\rho_0$  & $\chi^2_{r}$ &     $r_s$ & $\log\rho_0$  & $\chi^2_{r}$ \\ \hline
            A1 &      14.5 &       6.2 &       1.5 &      13.4 &       6.2 &       1.0 &      15.1 &       6.1 &       1.1 \\
            A2 &       4.7 &       8.6 &       0.6 &       4.5 &       8.7 &       0.5 &       6.4 &       8.2 &       1.1 \\
            A3 &       1.9 &       9.2 &       4.2 &       1.7 &       9.2 &       4.5 &       6.0 &       7.8 &       5.1 \\
            A4 &       7.9 &       7.5 &       2.0 &       6.6 &       7.6 &       1.6 &      17.9 &       6.7 &       1.3 \\
            A5 &       0.3 &      10.5 &       1.7 &       0.04 &      12.8 &      $> 10$ &       0.06 &      12.2 &       1.4 \\
            A6 &       6.6 &       7.9 &       1.4 &       5.8 &       8.0 &       1.4 &      23.4 &       6.7 &       1.4 \\
            A7 &       8.7 &       7.3 &       3.8 &       7.4 &       7.4 &       3.4 &      15.5 &       6.8 &       3.9 \\
            B1 & $\sim10^4$ &       2.9 &       6.2 & $\sim10^7$ &       1.2 &       1.4 & $\sim10^7$ &       0.08 &       $>10$ \\
            B2 &      18.6 &       6.2 &       4.5 &      14.8 &       6.1 &       3.2 &      15.6 &       6.1 &        $>10$ \\
            B3 &       3.9 &       8.4 &       2.4 &       3.7 &       8.4 &       2.4 &       4.5 &       8.2 &        $>10$ \\
            B4 & $\sim10^4$ &       3.3 &       2.9 &       5.0 &       7.4 &       6.6 & $\sim10^6$ &       1.8 &        $>10$ \\
            B5 &       9.0 &       7.4 &       1.8 &       7.7 &       7.5 &       2.1 &      15.1 &       6.8 &       $>10$ \\
            B6 &       5.3 &       8.2 &       7.2 &       2.0 &       9.0 &       $>10$ &      35.5 &       6.2 &        $>10$ \\
            B7 & $\sim10^4$ &       2.3 &       1.5 &       4.6 &       7.5 &       $>10$ & $\sim10^5$ &       1.7 &        $>10$ \\
    \end{tabular}
    }
  \caption{\footnotesize{NFW parameters for different galaxies and mass models with  same units as in given in Tab.\ref{tab:onlydm_bdm}}}
  \label{tab:onlydm_nfw}
\end{table}
\begin{equation}\label{eq:rhobdm}
\rho_{BDM}=\rho_0/[(r_c/r_s+r/r_s)(1+r/r_s)^2],
\end{equation}
which contains three parameters: a typical scale length ($r_s$) and density ($\rho_0$) of the halo, and a core radius ($r_c$). The BDM profile coincides with $\rho_{NFW}$ at large radius but has a core inner region, when the halo energy density $\rho_{BDM}$ reaches the value $\rho_c=E_c^4$ at  $r\simeq r_c$  with
\begin{equation}
\rho_{c} \equiv \rho_{BDM}(r=r_{c}) \simeq \rho_0 r_s/2 r_{c}.
\end{equation}
The relevant parameters to determine the inner core structure of galaxies are the value of $E_c=\rho_c^{1/4}$ and $r_c$. The parameters $\rho_0$ and $r_s$ depend on the initial conditions and formation of each galaxy.\\
\textbf{Analysis of LSB Galaxies.} We limit our sample to (early type and dwarf) THINGS galaxies with smooth, symmetric and
extended-to-large-radii rotation curves. The set consists of seventeen low luminous LSB galaxies. These observations represent the best available data to study the DM mass distribution and it has been used in works concerning the core versus cusp discrepancy
controversy \cite{deBlok:2008wp}. For technical details and systematic effects see \cite{Walter:2008wy}.
Our mass models include the three main components of a spiral galaxy: thin gaseous disk, $V_{gas}$, a thick stellar disk, $V_{\star}$, and a DM halo, $V_{halo}$. We have not considered a bulge because it is estimated to contain a small fraction of the total luminosity in our galaxy sample and therefore it has a small or n\emph{}ull effect in the analysis. The gravitational potential of the galaxy is the sum of each mass component, thus the observed rotational velocity is ${V_{obs}}^2 = {V^2_{halo}} + {V^2_{gas}} + {\gs V^2_{\star}}$.
Available photometry, extracted from the SINGS images at the $3.6 \mu m$, shows that the stars in our sample of galaxies are distributed in a thin disk, with exponential central surface density profile $\Sigma(R) = \Sigma_0 e^{-R/R_d}$. When $R_d$ is the scale length of the disk and $\Sigma_0$ is the central surface density with units [$M_\sun {\rm pc}^{-2}$]. The rotation velocity of an exponential disk is given by the well know Freeman formula. The contribution of the gaseous disk is directly derived from the HI surface density distribution.
The kinematics of stars brings a challenging problem in the analysis, mainly due to the uncertainty in the mass-to-light ratio ($\gs$), that is an
additional constant, free parameter in the mass model. The stellar contribution to rotational curves is in many cases dominant close to the galactic
center and this interferes with the determination of the core parameters. Some considerations has been made in order to reduce this uncertainty in
the parameters \cite{Kroupa:2000iv}, but still the stellar contribution is not well known and depends
on extinction, star formation history, initial mass function (IMF), among
other issues. We present a disk-halo decomposition using different assumptions for the stellar $\gs$: i) DM alone (Min.Disk); ii) DM and
gas (Min.Disk+gas); iii) Kroupa, considered as the minimal limit for the stellar disk; iv) diet-Salpeter, a stellar population synthesis model that
yields a maximum stellar disk; and v) Free $\gs$, treating the mass-to-light ratio as an extra free parameter in the model.\\
We have analyzed the rotation curves of seventeen galaxies using the above-mentioned five disk models and four different DM profiles (BDM, NFW, Burkert, ISO).  Since we have a large number of data, and because we want to emphasize the main results, here we present the general conclusions and we show results for  the Min.Disk scenario for BDM and NFW, only. However, a full comparison and a detailed analysis between all the different profiles  and mass models can be found in \cite{BDM2011}. We also perform 1 and  2 $\sigma$ likelihood contour plots for the BDM parameters, $r_c$ and $\rho_0$ for the different galaxies and mass models  \cite{BDM2011}.
We have grouped the galaxies into three blocks according to the ratio $r_c/r_s$ of the central fitted value for the min.disk scenario. The first group (G.A.) has a fitted $r_c/r_s < 1$, the second group (G.B.) has $r_c/r_s \simeq 1$, and finally, the third group (G.C.) has $r_c/r_s < 10^{-6}$. We shall explain the physical interpretation of the results for each group in the following paragraphs. Since the energy $E_c$ and $r_c$ can take values $(0, \infty)$ the correct distribution for the sample that we considered is the log-normal distribution, which is used to compute the statistics for the different BDM parameters.
\begin{figure}[!t]
  \subfloat{ \includegraphics[width=0.3\textwidth]{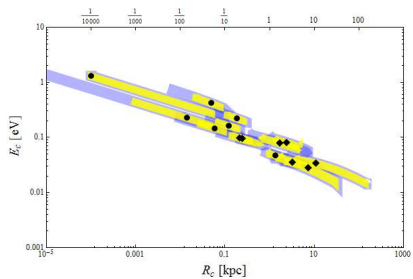}}
    \caption{\footnotesize{ We show  $r_c$ vs $E_c$ for Group A (circles ) and inner Group B (diamonds) galaxies, giving  $ <E_c>\simeq 0.1 {\rm \ eV}$ and core radius $<r_c> \simeq 260$ {\rm pc}. Yellow (light) and blue (dark) represent $\sigma$ and $2\sigma$ c.l. }}
 \label{fig:RcvsEcONLYDM}
\end{figure}
\begin{table}[h]
  \centering
  \scriptsize{
\begin{tabular}{r|r|rrr|rr}
           &            &     \multicolumn{ 3}{c|}{$\rho_{in}$} & \multicolumn{ 2}{c}{$\rho_{\alpha}$} \\
    Gal & $R_{\bf max}$ &     $r_c$ &  $\log \rho_0$ &     $r_s$ & $\log\rho_0$ &  $\alpha$ \\  \hline
    B1 &        3.9 &       3.27 &   4.40 &    1347.58 &       6.77 &          0 \\
    B2 &       2.27 &       7.40 &   7.81 &       3.80 &       7.45 &       0.08 \\
    B3 &        4.5 &       1.70 &   8.01 &       6.60 &        8.4 &       0.52 \\
    B4 &       0.97 &       0.25 &   5.38 &     569.39 &       7.95 &       0.84 \\
    B5 &       6.82 &       0.21 &   6.89 &      16.64 &       8.02 &        0.9 \\
    B6 &       3.73 &       2.42 &   8.38 &       5.58 &       8.54 &       0.24 \\
    B7 &       6.02 &      10.79 &   6.91 &      24.61 &       7.25 &       0.21 \\
\end{tabular}
    }
  \caption{\footnotesize{BDM parameters and the slope $\alpha$ obtained from the fittings of the inner galactic region with the $\rho_{in}$ and $\rho_{\alpha}$ profiles. The maximum distance is  $R_{\bf max}(kpc)$.}}
  \label{tab:bdm_inner}
\end{table}
We obtain seven galaxies (DDO 154 (A1), NGC 2841 (A2), NGC 3031 (A3), NGC 3621 (A4), NGC 4736 (A5), NGC 6946 (A6), and NGC 7793 (A7)) (we call Group A)
with fitted values $r_s > r_c \neq 0$ that are shown in  Table \ref{tab:onlydm_bdm}.
For these galaxies we obtain average values for:  the core $r_c \simeq 40$ {\rm pc}, $r_s \sim 5$ {\rm kpc}, and the energy $E_c \simeq 0.11 {\rm \ eV}$ for the min.disk analysis, that are typical values for a galaxy, except for DDO 154 which has a larger core, $r_c = 1.35$ {\rm kpc}.
 We notice that for all these galaxies the $\chi^2$ in BDM is smaller than NFW's but the extra parameter ($r_c$) in BDM makes the reduced $\chi^2_{red}$ equivalent for both profiles.
When comparing the BDM profile with the other two cored profiles, Burkert and ISO (Ref. \cite{BDM2011}) we conclude that BDM is still better. In particular, Burkert and ISO profiles have difficulties when fitting a couple galaxies (NGC4736, NGC 3621) having $\chi^2_{red} > 5$ while BDM and NFW each have $\chi^2_{red} \leq 2$.\\
In the second case, Group B, (IC 2574 (B1), NGC 2366 (B2), NGC 2903 (B3), NGC 2976 (B4), NGC 3198 (B5), NGC 3521 (B6), and NGC 925 (B7)), we get a fitted value of $r_c \simeq r_s$ and BDM profile fits better than NFW. In fact, the NFW  profile in some cases over-predicts the velocity in the inner parts of the galaxy or does fit with unphysical values for $r_s > \mathcal{O}(10^4)$ or $\rho_0 < \mathcal{O}(10^3) {\rm M_\sun/kpc^{3}}$, see Table \ref{tab:onlydm_bdm} and \ref{tab:onlydm_nfw}. The BDM profile gives $r_c \simeq r_s$ for all stellar mass models, having an average $r_c$ value of $\sim$6 and $\sim$5 {\rm kpc} for Min.Disk and Min.Disk+Gas, respectively. Since $r_c \simeq r_s$, implying that our BDM profile is of the type  $\rho \propto (r_c + r)^{-3}$ and hence having a strong core behavior.
For this set of galaxies the Burkert and ISO profiles, as well as the BDM, fit much better than NFW, indicating that a core is needed. The BDM and Burkert profiles are slightly better than ISO when we consider Min.Disk and Min.Disk+Gas analysis. Otherwise, for Kroupa and diet-Salpeter stellar mass models the BDM and ISO profiles have equivalent $\chi^2_{red}$ and fit slightly  better than Burkert.
To study in more detail the central region of the galaxies, we performed an analysis of the inner galactic region \cite{BDM2011}. We define the inner region as the set of data for each galaxy with a constant slope value for the observed velocity, and for radius smaller than the fitted $r_s$ in Table \ref{tab:onlydm_bdm},  and before the slope of $v$ starts to decrease \ci{BDM2011}. We computed the core radius, the central energy density, and the inner slope of the profile by examining the central region data with two different approaches: i) The first one by taking the limit $r \ll r_s$ in the BDM profile which gives $\rho_{in} = \rho_0 r_s (r_c+r)^{-1}$,  ii) The second is with the ansatz profile $\rho_{\alpha} = \rho_0 r^{-\alpha}$, where $\alpha$ is the  slope $\alpha\equiv -d\log[\rho]/\log[r]$.
Both approaches are related by the equation $0 \leq \alpha = r/(r_c + r) < 1$, from which we obtained different values for $\alpha$ depending on the analyzed region, i.e. the slope in BDM takes the values $\alpha=(0,1/2,1)$ for $r=(0,\rrc,\rrc\ll r\ll r_s)$. The results are in Table \ref{tab:bdm_inner} showing that in both approaches a core region is preferred and with a slope value in the interval $0 < \alpha \leq 0.52$ for most of the galaxies and a maximum slope of $\alpha \sim 0.84$, that implies that one is in the core dominant region, with $r/r_c=\alpha/(1-\alpha)$, or at most in the transition region between $r_c$ and $r_s$. We also obtain that in case i) the length of the core radius is reduced, the average $r_c = 1.8$ {\rm kpc} with energy $E_c \sim 0.06 {\rm \ eV}$ for min.disk  scenario, refer to Table \ref{tab:bdm_inner}. \\
We only present here the conclusions of  the last group C, composed of  three galaxies  (NGC2403, NGC50455
and NGC7331) \ci{BDM2011}. These galaxies have fitted values such that $r_c/r_s < 10^{-6}$ for most stellar mass models. The reduced number of data close to galactic center prevents us from finding whether these galaxies actually have or not a core. We conclude in \cite{BDM2011}
that ISO and Burkert are poorly fitting profiles, meanwhile BDM has a central value for $r_c \rightarrow 0$, so BDM reduces to NFW profile and both have equivalent fits. By applying inner analysis for these galaxies we obtain $0 \leq r_c \leq 50 $ {\rm pc} as a plausible interval that can fit the observations within the 2 $\sigma$ error, although a more definitive conclusion would demand more data close to the center in these galaxies.\\
We have also calculated fro the BDM profile the 1 and 2 $\sigma$ likelihood contour plots of $r_c$ and $\rho_0$ for the different galaxies and mass models, but due to lack of space here we refer to \ci{BDM2011}. In general we found that  when more mass components are included in the analysis the confidence level area becomes broader and in some cases the confidence levels increase up to an order of magnitude for the $r_c$ value \cite{BDM2011}.  Though  $r_c$ is different for each disk  mass model there is an interval of values where $r_c$ is consistent with all mass models within the 2 $\sigma$ errors. We also obtained that in the Free $\gs$ scenario, the fitted value never gets a greater value than diet-Salpeter, which is consistent with \cite{deBlok:2008wp}. We show in Fig. \ref{fig:RcvsEcONLYDM} the 1 and 2 $\sigma$ contour plots of $E_c$ vs  $r_c$  for the Min.Disk case and  in
Table \ref{tab:statistics_BDM} shows the averages for the different galaxies and mass models and the 1 $\sigma$ standard deviation of $E_c$ and $r_c$, for a detailed analysis see \cite{BDM2011}. We found that when more mass components are taking into account, mainly when the contribution of the stellar disk has a dominant behavior close to the galactic center, similiar as in Ref. \cite{Swaters09}, the evidence of the core fades away. We also found that even when $r_c$ is different for each mass model, its value lays inside the confidence levels obtained from the other mass models for the same galaxy.\\
\textbf{Energy Phase Transition $E_c$.} The values of the core radius $r_c$ and the transition energy $E_c=\rho_c^{1/4}$ are shown in
Table \ref{tab:onlydm_bdm}. In Fig. \ref{fig:RcvsEcONLYDM} we plot $E_{c}$ vs $r_c$ obtained from the minimal disk analysis
with its respectively contour errors. The circles and diamonds represent each galaxies from Group A and Group B, respectively. The energy of
transition between HDM and CDM takes place in Min.Disk mass model at energies and core radius up to 1 $\sigma$ c.l. :
\be\la{ec}
E_c=0.11 \times 10^{\pm 0.46}\; {\rm \ eV}, \hspace{.3cm} r_c = 260\times 10^{\pm 1.31}\; {\rm pc}.
\ee
The figures of $E_c$ vs $r_c$ for the other mass models  (Min.Disk+Gas, Kroupa, and diet-Salpeter mass models) are in \cite{BDM2011}.
Notice that the dispersion on $E_c$ is much smaller than that of $r_c$ and this is consistent with our BDM model
since $E_c$ is a new fundamental scale for DM model. Even though  we propose $E_c$  as a new fundamental constant for DM it is important to stress  that its value depends  on the choice of BDM profile and on high resolution data for the rotation curves.   We can also theoretically estimate  the value of $E_c$  and $\Lambda_c$ which are related via  $\rho_c\equiv E_c^4 = \Lambda_c\,n$.
We can then extract  information from the underlying gauge group using  gauge group dynamics. Using $\Lambda_c=\Lambda_i\,e^{-8\pi/b g^2_i}$ with $\Lambda_c\simeq E_c= 0.1\,{\rm \ eV}$ and $g^2_i=1/2$ the gauge coupling at unification or inflation scale $ \Lambda_i=10^{16}$ GeV.
We obtain   $b=8\pi/(g^2_iLog[\Lambda_i/\Lambda_c])\simeq 1$ and using that for a SUSY  gauge group one has  $b=3N_c-N_f$ we could
have for example a gauge group $SU(N_c=3)$ with $N_f=8$ fundamental particles.
The energy transition in Eq.(\ref{ec}) is similar to the mean energy of a relativistic neutrino which is such that $\sum m_\nu < 0.58 {\rm \ eV} (95\% {\rm \ CL)}$ if one assumes 3.04 neutrino species with degenerate mass eigenstates.  Furthermore, an interesting connection could
be further explored  between the phase transition scale $E_c$ of our BDM model, which also sets the mass of the BDM particles, to the neutrino mass generation mechanism \ci{NBDM}. However, we would like to emphasize that our BDM are not neutrinos, since neutrinos
are HDM,  and a combination of CDM plus neutrinos would have a cuspy NFW profile with CDM dominating in the inner region of the galaxies \cite{Navarro:1995iw}.
A main difference between neutrinos and BDM is that neutrinos were in thermal equilibrium at $E\gg$ MeV with the standard model (SM) particles but BDM may not have been  in thermal equilibrium with SM, as for example axion fields which are CDM but have a small mass, $m<$ eV.
\begin{table}[h]
  \scriptsize{
    \begin{tabular}{l|r|rrr|r|r|rrr|r}
                                                              \multicolumn{ 11}{c}{BDM Statistics} \\
               &                \multicolumn{ 5}{c|}{Energy $E_c$} &                  \multicolumn{ 5}{c}{Core $r_c$} \\
   Mass Models & $\tilde{E}_{c_n}$ & $E_{c_{-}}$ & $\hat{E}_c$ & $E_{c_{+}}$ & $\sigma_{E_c}$ &  $\tilde{r}_{c_n}$ & $r_{c_{-}}$ & $\hat{r}_c$ & $r_{c_{+}}$ & $\sigma_{r_c}$ \\
               \hline
      Min.Disk &       0.14 &       0.04 &       0.11 &       0.33 &  0.46   &  1.91 &       0.01 &       0.26 &       5.41 & 1.31\\
  M.Disk+gas &       0.09 &       0.05 &       0.08 &       0.16 &   0.27  & 1.13 &       0.12 &       0.63 &       3.26 & 1.31  \\
        Kroupa &       0.09 &       0.02 &       0.06 &       0.21 &  0.52   &6.92 &       0.13 &       1.67 &      21.59 &  1.11\\
 diet-Salpeter &       0.03 &       0.02 &       0.03 &       0.05 &  0.24   &18.08 &       3.22 &      12.27 &      46.68 &  0.58\\
    \end{tabular}
  \caption{\footnotesize{We show the statistics  $E_c(eV)$ and  $r_c(kpc)$ for the BDM profile for the different mass models. In Columns (2) is the arithmetic mean energy, $E_{c_n}(eV)$, in columns (3-5) we present the median energy $\hat{E}_c$ and  $E_{c \pm} \equiv \hat{E}_c 10^{\pm \sigma}$ assuming a log-normal distribution.  In columns (6-9) we present the same statistics for $r_c(kpc)$. }}
  \label{tab:statistics_BDM}
  }
\end{table}
\\ \textbf{Conclusions.} To summarize, we have tested different DM profiles with several mass models using LSB THINGS galaxies.  The study
favors BDM (core profiles) over the cuspy NFW, in accordance to general results found in other works \cite{deBlok:2008wp,Salucci:2007tm,deBlok10}. However  different systematics may
play an important role in the observations such as noncircular motions, resolution of data and other issues \cite{vandenBosch:1999ka, deBlok:2008wp}.  We presented here the analysis with Min.Disk for BDM and NFW profiles, and further we have performed the study using the five different disk mass components
for the BDM, NFW, Burkert, and Pseudo-Isothermal profiles in \ci{BDM2011}.
We  performed an analysis of the inner regions of  Group B and C galaxies showing the ability to extract information  from the rotation curves and the results are consistent with a core profile and we compute the core radius, central density and inner slope corresponding to the core region. In the cases where $r_c/r_s < 10^{-6}$ the BDM is indistinguishable from NFW profile and much better than Burkert or  ISO profiles. The likelihood contour plots reveal how the central contribution of the stellar disk fades away the evidence of the core making $r_c \rightarrow 0$. The observational resolution of the THINGS sample is of high quality, but one still needs data closer to the center of some galaxies, on scales smaller than 200 {\rm pc}, in order to discern between cored or cuspy profiles.  This is because stars pose a very challenging problem when testing the core-cusp problem, largely due to the uncertainty of the mass-to-light ratio and their dominant behavior close to the center of the galaxy.
We computed the 1 and 2 $\sigma$ confidence levels for the BDM parameters  $E_c$ and $r_c$ for the different galaxies and mass models. The contours plots show consistency in the fitted values of $E_c$ and $r_c$ for the different mass models for each galaxy. We found that the energy of transition between HDM and CDM takes places in Min.Disk mass models with an average energy $E_c = 0.11^{+0.21}_{-0.07} {\rm \ eV}$ and with a core radius $r_c \sim 260 \times 10^{\pm 1.31}\, {\rm pc}$, in accordance with Ref. \cite{de Blok:1996ns}.  Confidence level contours show that the core radius depends on the number of disk mass components taking into account in the analysis.  We found that $r_c$ is highly constrained if the stellar disk has a dominant behavior close to the center of the galaxy.  However, we notice that even though $r_c$ is different for each disk  mass model there is an interval of values where $r_c$ is consistent for all mass models within the 2 $\sigma$ errors. The dispersion on $E_c$ is much smaller than that of $r_c$
and this is consistent with our BDM model since $E_c$ is a new fundamental scale for DM model while $r_c$ depends on $r_s,\rho_0$ which depends on the formation and initial conditions for each galaxy. Given the richer structure that the BDM profile has, through its transition from CDM to HDM (when $\rho \simeq \rho_c$),  it allows for a physical explanation of the rotation curves of the different galaxies.

\acknowledgments
We thank E. de Blok for providing the observational data of THINGS and C. Frigerio for very useful help. A.M. and J.M.  acknowledge financial support from  Conacyt Proyect 80519 and J.L.C.C. from Conacyt Proyect  84133-F.

\thebibliography{}
\footnotesize{
\bibitem{KrAl78} P. C. van der Kruit and R. J. Allen,  Annu. Rev. Astron. Astrophys. \textbf{16}, 103 (1978)
V. Trimble,  Annu. Rev. Astron. Astrophys.   \textbf{25}, 425 (1987)
Y. Sofue and V. Rubin,  Annu. Rev. Astron. Astrphys.   \textbf{39}, 137 (2001)
\bibitem{deBlok10}
W.J.G. de Blok,
Advances in  Astronomy  {\bf 2010},  Article ID 789293 (2010), arXiv:0910.3538

\bibitem{Navarro:1995iw}
  J.~F.~Navarro, C.~S.~Frenk and S.~D.~M.~White,
  Astrophys.\ J.\  {\bf 462}, 563 (1996)
\bibitem{Bu95} A. Burkert,  Astrophys. J.   \textbf{447}, L25 (1995)

\bibitem{deBlok:2008wp}
  W.~J.~G.~de Blok, F.~Walter, E.~Brinks, C.~Trachternach, S.~H.~Oh and R.~C.Kennicutt,
  Astrophys.\ J.\  {\bf 136}, 2648 (2008)
 
 \bibitem{Gentile:2004tb}
  G.~Gentile, P.~Salucci, U.~Klein, D.~Vergani and P.~Kalberla,
  Mon.\ Not.\ Roy.\ Astron.\ Soc.\  {\bf 351}, 903 (2004)
 F.~Donato, G.~Gentile, P.~Salucci, C.~F.~Martins, M.~I.~Wilkinson, G.~Gilmore, E.~K.~Grebel, A.~Koch {\it et al.},  Mon.\ Not.\ Roy.\ Astron.\ Soc.\  {\bf 397}, 1169 (2009)

\bibitem{KuMcBl08}
 R. Kuzio de Naray, S. S. McGaugh, and W.J.G. de Blok, Astrophys.  J.  {\bf 676}, 920 (2008)
\bibitem{vandenBosch:1999ka}
  F.~C.~van den Bosch, B.~E.~Robertson, J.~J.~Dalcanton and W.~J.~G.~de Blok,
  Astron.\ J.\  {\bf 119}, 1579 (2000)

\bibitem{EvAnWa09} N. W. Evans, J. An, and M. G. Walker,
Mon. Not. R. Astron. Soc. {\bf 393}, L50(2009).

\bibitem{SpGiHa05}
K. Spekkens, R. Giovanelli, M. P. Haynes,  Astron. J.   \textbf{129}, 2119 (2005)

   \bibitem{Pontzen:2011}
  A. Pontzen, F. Governato,
  ``How Supernova Feedback Turns Dark Matter Cusps Into Cores''
  arXiv:1106.0499 [astro-ph.CO]
  G.~Ogiya and M.~Mori,
  arXiv:1106.2864 [astro-ph.CO]
\bibitem{delaMacorra:2009yb}
  A. de la Macorra,
 Astropart.\ Phys.\  {\bf 33}, 195 (2010)
 \bibitem{BDM2011}
J.~ Mastache, A.~ de la Macorra, and J. L. Cervantes-Cota,
arXiv:1107.5560 [astro-ph.CO]

\bibitem{Walter:2008wy}
  F.Walter {\it et al.},
  Astron.\ J.\  {\bf 136}, 2563 (2008)
 arXiv:0810.2125

\bibitem{Macorra.DE}
  A.~De la Macorra  and C.~Stephan-Otto,
  Physical Review Letters 87, 271301 (2001)
  JHEP {\bf 0301}, 033 (2003)
  ;  A.~de la Macorra,
  Phys.\ Rev.\  D {\bf 72}, 043508 (2005).
\bibitem{Macorra.DEDM}
 A.~de la Macorra,
  Phys.\ Lett.\  B {\bf 585}, 17 (2004).

 \bibitem{Kroupa:2000iv}
  P.~Kroupa,
  Mon.\ Not.\ Roy.\ Astron.\ Soc.\  {\bf 322}, 231 (2001).

 \bibitem{Swaters09}
 R.A. Swaters, R. Sancisi, T.S. van Albada, J.M. van der Hulst
 Astron. \& Astrophys.   {\bf 493}, 871 (2009);
 ; ibid  Astrophys.\ J.\  {\bf 729}, 118 (2011).
 \bibitem{NBDM}
A.~ de la Macorra and J.~ Mastache, in preparation

\bibitem{Salucci:2007tm}
  P.~Salucci, A.~Lapi, C.~Tonini, G.~Gentile, I.~Yegorova and U.~Klein,
  Mon.\ Not.\ Roy.\ Astron.\ Soc.\  {\bf 378}, 41 (2007)
  [arXiv:astro-ph/0703115];
 P.~Salucci,
 arXiv:1008.4344

\bibitem{de Blok:1996ns}
  W.~J.~G.~de Blok, S.~S.~McGaugh and J.~M.~van der Hulst,
  Mon.\ Not.\ Roy.\ Astron.\ Soc.\  {\bf 283}, 18 (1996).
  W.~J.~G.~de Blok, S.~S.~McGaugh and V.~C.~Rubin,
  Astron.\ J.\  {\bf 122}, 2396 (2001)

}

\end{document}